\documentstyle[11pt,epsf,newpasp,twoside]{article}
\def\HST{{\sl HST\/ }}
\def\kms{\nobreak\mbox{$\;$km\,s$^{-1}$}}
\def\mag{\hbox{$.\!\!^{\rm m}$}}

\begin{document}

\title{The Status of the Distance Scale \\ 
{\rm \normalsize (A Report on the ``{\sl HST\/} SNe\,Ia Calibration
  Program for the Hubble Constant'' by A.~Sandage, A. Saha,
  L. Labhardt, N. Panagia, F.D. Macchetto, and G.A.~Tammann)}}

\author{G.A. Tammann, B.R. Parodi, \& B. Reindl}

\affil{Astronomisches Institut der Universit\"at Basel, \\
       Venusstr.~7, CH-4102 Binningen, Switzerland}

\begin{abstract}
Cepheids are the pillar of the extragalactic distance scale, but their
reach in distance is not sufficient to calibrate $H_0$. Yet {\sl
  HST\/} has provided Cepheid distances to eight galaxies which have
produced SNe\,Ia. The latter are used as nearly perfect standard candles
to carry the distance scale to $30\,000\kms$, giving
$H_0=58\pm5$. This determination finds support in other methods but
its range in distance and its weight is unparalleled.
\end{abstract}

\section{Introduction}
The determination of large scale distances and hence the calibration
of the Hubble constant $H_0$ has been fundamentally influenced by the
'Hubble Space Telescope' ({\sl HST}).
For the first time this extremely powerful instrument has made
accessible a simple, reliable, and clear-cut method to reach into the
{\em cosmic\/} expansion field by calibrating the luminosity of type
Ia supernovae (SNe\,Ia) -- the best standard candles known -- through
Cepheid distances of their nearest representatives. The dominant
r{\^o}le of Cepheids on this route to $H_0$ has presumably prompted
the organizers of this Colloquium to ask for a status report on the
(distant) distance scale. The Cepheid surveys to be reported here
extend only tiny angles in the sky, but they comprise large parts of
or even entire galaxies and in this sense may also justify the
Colloquium's emphasis on ``Large-Scale Surveys''. 

In Section~2 we discuss Cepheids as distance indicators and quote the
Cepheid distances of eight galaxies which have produced nine SNe\,Ia.
The Cepheid-calibrated SNe\,Ia luminosities are given in Section~3. In
Section~4 we discuss the Hubble diagram of an objective sample of blue
SNe\,Ia and homogenize the SNe\,Ia as to decline rate $\Delta m_{15}$
and colour $(B-V)$. The data in Sections~3 and 4 are combined in
Section~5 to derive the large-scale value of $H_0$.
The result is compared with external evidence in Section~6. Some
conclusions follow in Section~7.

\section{Cepheids as Distance Indicators}
Cepheids are presently, through their period-luminosity (PL)
relation, the most reliable and least controversial distance
indicators. The slope and the zeropoint of the PL relation is taken
from the very well observed Cepheids in the Large Magellanic Cloud
(LMC), whose distance modulus is adopted to be $(m-M) = 18.50$
(Madore \& Freedman 1991).

   An old PL relation calibrated by Galactic Cepheids in open clusters,
and now vindicated by Hipparcos data (Sandage \& Tammann 1998), gave
$(m-M)_{\rm LMC}=18.59$ (Sandage \& Tammann 1968, 1971). Hipparcos data
combined with more modern Cepheid data give an even somewhat higher modulus
(Feast \& Catchpole 1997). Reviews of Cepheid distances (Federspiel,
Tammann, \& Sandage 1998; Gratton 1998; Feast 1999; Walker 1999a)
cluster around $18.56\pm0.05$, 
-- a value in perfect agreement with the purely geometrical distance
determination of SN\,1987A ($18.58\pm0.05$; Gilmozzi \& Panagia
1999; Sonneborn et~al.\ 1997; Walker 1999a). 
From RR~Lyr stars Gratton (1998) concludes that $(m-M)_{\rm
  LMC}=18.54\pm0.12$.

   A general problem of distance determinations is that distance
indicators are occasionally used which have never been demonstrated to
be useful and reliable. Their application relies on the {\em
  assumption\/} of uniformity. Examples are statistical parallaxes of Galactic
RR~Lyr stars, which are sensitive to the sample selection (Walker
1999b), and red-giant clump stars, whose luminosity is expected to be
metal- and age-dependent (Chaboyer 1999). These untested methods yield
typically small LMC distances and have led to what has recently been
termed the ``short distance scale''. The corresponding LMC distances
carry low weight.

   There has been much debate about the possibility that the PL
relation of Cepheids depends on metallicity. The question here is
more of principal than of practical importance because the {\em mean\/}
metallicity of the seven spiral galaxies and one Am galaxy considered
below hardly differs by much from the Galactic metallicity. Direct
{\em observational\/} evidence for a (very) {\em weak\/} metallicity
dependence comes from the fact that the metal-rich
Galactic Cepheids give perfectly reasonable distances for both the
moderately metal-poor LMC Cepheids and the really metal-poor SMC
Cepheids and, still more importantly, that their relative distances
are wavelength-independent (Di Benedetto 1997; cf. Tammann 1997).~-- 
Much progress has been made on the theoretical front. Saio
\& Gautschy (1998) and Baraffe et~al.\ (1998) have evolved Cepheids
through the different crossings of the instability strip and have
investigated the pulsational behavior at any point. The resulting
(highly metal-insensitive) PL relations in bolometric light have been
transformed into PL relations at different wavelengths by means of
detailed atmospheric models; the conclusion is that any metallicity
dependence of the PL relations is negligible (Sandage, Bell, \&
Tripicco 1998; Alibert et~al.\ 1999; cf. however Bono, Marconi, \&
Stellingwerf 1998, who strongly depend on the treatment of stellar
convection). 

   While remaining uncertainties of the PL relation and the zeropoint
seem to have only minor practical consequences, the application to
{\sl HST\/} observations is by no means simple. The photometric zeropoint,
the linearity over the field, crowding, and cosmic rays raise
technical problems. The quality of the derived distances depends
further on (variable) internal absorption and the number of available
Cepheids in view of the finite width of the instability strip. (An
attempt to beat the latter problem by using a PL-color relation is
invalid because the underlying assumption of constant slope of the
constant-period lines is unjustified; cf. Saio \& Gautschy
1998). Typical errors of individual Cepheid distances from \HST are
therefore $\pm0.2\,$mag (10\% in distance). For four of the nine
SNe\,Ia in Table~1 the resulting errors in luminosity are
smaller, because they suffer closely the same (small) absorption as
``their'' Cepheids, such that only apparent distance moduli are needed.

   The nine SNe\,Ia in the eight galaxies with known Cepheid distances
from {\sl HST\/} are listed in Table~1. Six galaxies are
from the ``{\sl HST\/} SNe\,Ia Calibration Program for the Hubble
Constant''; two additional ones are taken from the literature. The
distance moduli are based on an adopted, conservative value of
$(m-M)_{\rm LMC}=18.50$.

%
\begin{table}[t]
\begin{center}
\scriptsize \rm
\caption{Absolute $B$, $V$,  and $I$ magnitudes of blue SNe\,Ia calibrated
  through Cepheid distances of their parent galaxies. (Errors in
  $0\mag01$ in parentheses)}
\label{tab:1}
\hspace*{-2.5cm}
\begin{minipage}{\textwidth}
\begin{tabular}{llclcccccc}
\noalign{\smallskip}
\hline
\noalign{\smallskip}
  \multicolumn{1}{c}{SN} & \multicolumn{1}{c}{Galaxy} & 
  \multicolumn{1}{c}{$\log v$} & 
  \multicolumn{1}{c}{(m-M)$_{\rm AB}$} & (m-M)$_{\rm AV}$ &
  ref. &
  M$^0_{\rm B}$ & M$^0_{\rm V}$ & M$^0_{\rm I}$ & 
  $\Delta m_{15}$ \\[2pt]
  \multicolumn{1}{c}{(1)} & \multicolumn{1}{c}{(2)} &
  \multicolumn{1}{c}{(3)} & \multicolumn{1}{c}{(4)} &
  \multicolumn{1}{c}{(5)} & \multicolumn{1}{c}{(6)} &
  \multicolumn{1}{c}{(7)} & \multicolumn{1}{c}{(8)} &
  \multicolumn{1}{c}{(9)} & \multicolumn{1}{c}{(10)} \\
\noalign{\smallskip}
\hline
\noalign{\smallskip}
1895B  & NGC\,5253  & 2.464 & 28.13\,(08)     & 28.10\,(07) &
 2 & -19.87\,(22) & $\cdots$     & $\cdots$     & $\cdots$   \\  
1937C  & IC\,4182   & 2.519 & 28.36\,(09)     & 28.36\,(12) &
 1 & -19.56\,(15) & -19.54\,(17) & $\cdots$     & 0.87\,(10) \\
1960F  & NGC\,4496A & 3.072 & 31.16\,(10)     & 31.13\,(10) &
 3 & -19.56\,(18) & -19.62\,(22) & $\cdots$     & 1.06\,(12) \\
1972E  & NGC\,5253  & 2.464 & 28.13\,(08)     & 28.10\,(07) &
 2 & -19.64\,(16) & -19.61\,(17) & -19.27\,(20) & 0.87\,(10) \\
1974G  & NGC\,4414  & 2.820 & 31.46\,(17)$^a$ & $\cdots$   &  
 4 & -19.67\,(34) & -19.69\,(27) & $\cdots$     & 1.11\,(06) \\
1981B  & NGC\,4536  & 3.072 & 31.10\,(05)$^a$ & $\cdots$   & 
 5 & -19.50\,(14) & -19.50\,(10) & $\cdots$     & 1.10\,(07) \\
1989B  & NGC\,3627  & 2.734 & 30.22\,(12)$^a$ & $\cdots$    & 
 6 & -19.47\,(18) & -19.42\,(16) & -19.21\,(14) & 1.31\,(07) \\
1990N  & NGC\,4639  & 3.072 & 32.03\,(22)$^a$ & $\cdots$    & 
 7 & -19.39\,(26) & -19.41\,(24) & -19.14\,(23) & 1.03\,(05) \\
1998bu & NGC\,3368  & 2.814 & 30.37\,(16)$^a$ & $\cdots$    & 
 8 & -19.76\,(31) & -19.69\,(26) & -19.43\,(21) & 1.08\,(05) \\
\noalign{\smallskip}
\hline
\noalign{\smallskip}
 & & \multicolumn{4}{l}{mean (straight, excl. SN\,1895B)} & 
     -19.57\,(04) & -19.56\,(04) & -19.26\,(06) & \\ 
 & & \multicolumn{4}{l}{mean (weighted, excl. SN\,1895B)} & 
     -19.55\,(07) & -19.53\,(06) & -19.25\,(09) & \\
\noalign{\smallskip}
\hline
\noalign{\smallskip}
\end{tabular}
\end{minipage}
\begin{minipage}{0.9\textwidth}
{\footnotesize 
$^a$\,The {\em true\/} distance modulus is listed.\\[5pt]
{\sc References:} 
(1) Saha et~al.\ 1994; 
(2) Saha et~al.\ 1995;
(3) Saha et~al.\ 1996b;
(4) Turner et~al.\ 1998;
(5) Saha et~al.\ 1996a;
(6) Saha et~al.\ 1999;
(7) Saha et~al.\ 1997;
(8) Tanvir et~al.\ 1995.
}
\end{minipage}
\end{center}
\end{table}

The Cepheids observed with the wide-field camera of {\sl HST\/} (WFP2)
are corrected by $+0\mag05$ for the photometric short exposure/long
exposure zeropoint effect of that camera (Stetson et~al.\ 1998; Saha
et~al.\ 1995).

   The Cepheid distances in Table~1 have been re-analysed by
Gibson et~al.\ (1999). While their object to object photometry
corroborates  the original sources to within a few percent, their
treatment of Cepheid reddening and sample selection is objectionable,
leading to distance moduli $0\mag13$ smaller on average. The prize is
that their resulting SNe\,Ia luminosities have a noticeably larger
scatter ($\sigma_{\rm B}=0\mag16$ instead of $\sigma_{\rm
  B}=0\mag12$). Yet the most important point is that their adopted
{\em mean\/} SNe\,Ia luminosities are fainter than those in
Table~1 by only $0\mag05$, $0\mag07$, and $0\mag15$ (the
latter value from only four SNe\,Ia) in $B$, $V$, and $I$,
respectively. These minute differences are even below the statistical
errors.

   On the other hand a re-analysis of NGC\,4414 (Turner et~al.\ 1998) 
with an independent photometry package (ROMAPHOT) suggests that the $B$
magnitudes were measured $\sim\!0\mag10$ too bright (Thim 1999). 

   The most severe problem of the Cepheid distances is selection
bias. The intrinsic width of the instability strip poses the problem
that at any given period a fair number of bright and faint Cepheids
must be sampled. But at shorter periods the faint counterparts are
progressively discriminated. This leads to a systematic {\em
  underestimate\/} of the distance (Sandage 1988). The effect becomes
apparent if sufficient numbers of Cepheids are available because the
derived distance increases as one goes to longer periods. Narasimha \&
Mazumdar (1998) and Mazumdar (1999) suggest that in the case of
NGC\,4321 (M\,100) (Ferrarese et~al.\ 1996) the distance was
underestimated by $\sim\!15\%$ due to bias. The distance moduli in
Table~1 were derived with the Sandage bias in mind, but too
few Cepheids are available for a rigid treatment. In principle the
bias problem remains that Cepheid distances tend to be underestimates
if one works near the detection limit. The severity of the bias
decreases as the available period interval increases.

\section{The Calibration of SNe\,Ia Luminosities from Cepheid
  Distances}
The absolute $B$, $V$, $I$ magnitudes at maximum of the nine SNe\,Ia
in galaxies with Cepheid distances are shown in Table~1. The
adopted apparent maximum magnitudes of the SNe\,Ia as well as the
internal absorption of the last five entries in Table~1 are
detailed elsewhere (Parodi et~al.\ 1999). The straight and weighted
mean of the eight adopted SNe\,Ia (SN\,1895B is excluded because it
has no $M_{\rm V}(\max)$) is also shown in Table~1.

The r.m.s. deviation of a single $M$-value in Table~1
amounts to only $\sigma_{\rm B}=\sigma_{\rm V} =0\mag12$ confirming
the notion of blue SNe\,Ia being exceptionally useful standard
candles. In fact, the deviations are smaller than the estimated errors
of the Cepheid distances. This means that the Cepheid distances
are better than expected, and there is little room for distance errors
due to metallicity differences.

   The empirical luminosity calibration is in perfect agreement with
presently available theoretical models. H\"oflich \& Khokhlov (1996)
have fitted sufficiently blue models, i.e. $(B-V)<0.2$ at maximum, to
the light curves and spectra of 16 SNe\,Ia. Their mean luminosity in
$B$ and $V$ is only $0\mag05$ fainter than in Table~1 . In a
recent review Branch (1998) 
has concluded that present theory is best satisfied by $M_{\rm B}
\approx M_{\rm V}=-19.4$ to $-19.5$ for blue SNe\,Ia.

   Della Valle et~al.\ (1998) have attempted to derive an independent
distance of NGC 1380, host of SN\,1992A, by means of the peak of
the luminosity function of globular clusters (GCLF) and advocated a
low luminosity of the SN. However, the GCLF method is known to give
sometimes erratic results (Tammann 1998b).

   Kennicutt, Mould, \& Freedman (1998), and Freedman (1999) have
discarded several of the calibrators in Table~1 and added two that are
not based on direct Cepheid distances to the host galaxy. They have
consequently derived a fainter mean absolute $B$ magnitude than
in Table~1. Specifically, they assume that the distance of
the early-type galaxies NGC\,1316 and NGC\,1380 in the Fornax cluster,
parent galaxies of SN\,1980N and SN\,1992A, are identical with that of
the spiral NGC\,1365 for which a Cepheid distance is available. Suntzeff
et~al.\ (1999) have also considered the questionable SN\,1980N and
SN\,1992A as possible calibrators. However, with the significantly
larger Cepheid distance of an other Fornax spiral (NGC\,1425; Mould
et~al.\ 1999) this has become untenable.

    The \HST project for the luminosity calibration of SNe\,Ia will be
continued. Cepheid observations are presently granted for NGC\,4527
(with the peculiar-spectrum and possibly overluminous SN\,1991T)
and NGC\,3982 (with SN\,1998aq).

\section{The Hubble Diagram of Distant SNe\,Ia}
A complete sample of 45 SNe\,Ia has been collected from the
literature, the majority from the Cal{\'a}n/Tololo survey (Hamuy
et~al.\ 1996). They fulfill the condition that their values $m_{\rm
  B}(\max)$ and $m_{\rm V}(\max)$ are available; wherever $m_{\rm
  I}(\max)$ is known it is retained. They must have occurred after
1985 to ensure the quality of their photometry, and they must have
recession velocities of $v_{220}>1000\kms$ (after correction for
Virgocentric streaming velocities) to guard against excessive effects
of peculiar velocities. Finally they must be {\em blue},
i.e. $(B_{\max}-V_{\max})\le0.20$; this is to exclude peculiar SNe\,Ia,
which are known to be red and underluminous, as well as SNe\,Ia which
suffer much absorption in their parent galaxies. All SNe\,Ia are
corrected for Galactic absorption (Schlegel, Finkbeiner, \& Davis 1998).

   The SNe\,Ia of the sample, for which spectra are known, are all
``Branch normal'' (Branch, Fisher, \& Nugent 1993). The only exceptions
are SN\,1991T and 1995ac; they are probably overluminous and excluded
here. Only seven SNe\,Ia of the sample have $(B-V)>0.06$. They are
underluminous judging from their velocity distance, and five of them 
lie close to the centre of their parent spiral galaxies. Their redness is
probably due to internal absorption. They are left out in the
following. Had they been retained in on the assumption that their color is
intrinsic, the derived value of $H_0$ would become somewhat {\em
  lower}. If they were retained and corrected for standard absorption
they would have an unnoticeable effect on $H_0$ as derived below.

   The remaining sample of 36 SNe\,Ia has a small range in color of
$-0.10\le(B-V)\le0.06$ and a mean color of $<\!\!B-V\!\!>\;=
-0.010\pm0.09$. The mean color of the SNe\,Ia in E/S0 galaxies, of the
outlying SNe\,Ia in spirals, and of the calibrators in
Table~1 -- which are expected to suffer minimum internal
absorption or are corrected for absorption -- is  $<\!\!B-V\!\!>\;=
-0.014\pm0.011$. The close agreement of these two mean colors is a
convincing demonstration that the sample of 36 SNe\,Ia is essentially
free of internal absorption. This is not surprising because unreddened
SNe\,Ia have at large distances a considerably higher discovery
chance.

   The 36 SNe\,Ia define tight Hubble diagrams in $B$, $V$, and
$I$. The scatter is only $\sigma=0.23$, $0.22$, and $0.18$,
respectively. 

   Blue SNe\,Ia at maximum phase still show some variation in
temperature, spectrum, and presumably Ni mass. It is hence indicated
to ask whether these parameters correlate with luminosity. However,
data on the relevant parameters are too sparse to establish a firm
relation. Instead it is reasonable to seek for empirical correlations
between luminosity and directly observable ``second parameters'' like
decline rate $\Delta m_{15}$ (i.e. the luminosity decrease in $B$
magnitudes within 15 days after $B$ maximum), SN colour, Hubble type
of the parent galaxy, and radial distance from the centre of the
parent galaxy.

   In fact a luminosity dependence on $\Delta m_{15}$, SN color
$(B-V)$, and Hubble type $T$ does exist. However, an increase of
the luminosity scatter with decreasing galactocentric distance
(Wang, H{\"o}flich, \& Wheeler 1997) is not confirmed by blue SNe\,Ia.

   To obtain quantitative relations between the maximum magnitudes and
any of the second parameters, one must correlate the residua (read in
magnitude) from the mean Hubble line with slope 0.2 with $\Delta
m_{15}$, $(B-V)$, and $T$. However, there are strong reasons to
believe that the residua of SNe\,Ia inside $10\,000\kms$ are not only
caused by luminosity variations, but also by deviation from a pure Hubble 
flow. If only the SNe\,Ia with $v>10\,000\kms$ are considered one finds
\begin{eqnarray}
  \label{eq:1}
  B^{\rm corr} & = & B - 0.34(\Delta m_{15} - 1.1) - 2.45(B-V), \\
  V^{\rm corr} & = & V - 0.41(\Delta m_{15} - 1.1) - 1.41(B-V), \\
  I^{\rm corr} & = & I\; - 0.43(\Delta m_{15} - 1.1) - 1.37(B-V),
\end{eqnarray}
When these equations are applied to the 36 SNe\,Ia of the sample one
obtains homogenized magnitudes $m^{\rm corr}$ as if all SNe\,Ia had
$\Delta m_{15} = 1.10$ and $(B-V) = 0.00$. The second parameters
$\Delta m_{15}$ and $(B-V)$ are nearly orthogonal to each other, but
the corrected magnitudes show no remaining dependence on Hubble type
$T$. (SNe\,Ia in E/S0 galaxies are on average fainter than those in
spirals, but they are also faster decliners).

   The 36 corrected SNe\,Ia define Hubble diagrams in $B$, $V$, and
$I$ as shown in Fig.~1. A fit to the SNeIa with $v>10\,000\kms$ gives:
\begin{eqnarray}
\label{eq:4}
  \log v & = & 0.2\,m_{\rm B}^{\rm corr} + (0.660\pm0.006); 
           \quad \sigma_{\rm B}=0.13, \quad N=19 \\
  \log v & = & 0.2\,m_{\rm V}^{\rm corr} + (0.663\pm0.006); 
           \quad \sigma_{\rm V}=0.13, \quad N=19 \\
  \log v & = & 0.2\,m_{\rm I}^{\rm corr} + (0.610\pm0.007);
           \quad \sigma_{\rm I}\;=0.13, \quad N=14. 
\end{eqnarray}
\begin{figure}
\epsfxsize=9cm   
\hspace*{1.9cm}\epsfbox{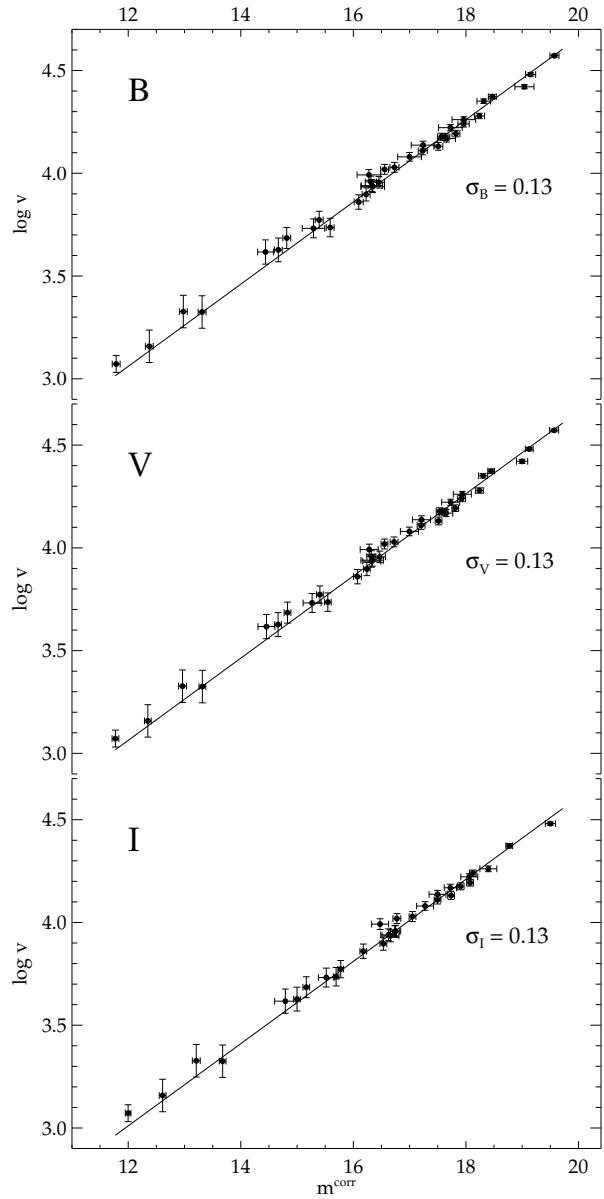} 
\caption{The Hubble diagram in $B$, $V$, and $I$ for 36 blue
  SNe\,Ia.  The fitted lines have slope 0.2; they
  only consider SNe\,Ia with $v>10\,000\kms$. -- The velocities are
  corrected for a self-consistent Virgocentric infall model
  (Kraan-Korteweg 1986) if $v<3000\kms$; for $v>3000\kms$ an
  additional correction was applied for the motion of $630\kms$
  relative to the CMB dipole anisotropy (Smoot et~al.\ 1992). -- (From
  Parodi et~al.\ 1999).}
\label{fig:1}
\end{figure}

   The scatter in magnitude of $\sigma = 0.13$ is now impressively
small, in fact it could be caused entirely by observational
error. The intrinsic luminosity scatter of homogenized SNe\,Ia is
therefore below the present detection limit.

   The scatter of the nearer SNe\,Ia with $v<10\,000\kms$ is somewhat
larger, i.e. $\sigma = 0.16$. This must be due to peculiar motions in
the order of $\Delta v/v \approx 0.05$, which is most reasonable
considering our local motion of $630\kms$ with respect to the
CMB. There is also a suggestion of the nearer SNe\,Ia to lie somewhat
above the mean Hubble line, which seems to suggest  a slightly
higher local value of $H_0$ (see below).

\section{{\boldmath $H_0$} from SNe\,Ia}
Simple transformation of equations~(4$-$6) leads to
\begin{eqnarray}
  \label{eq:7}
  \log H_0 & = & 0.2\,M^{\rm corr}_{\rm B} + (5.660 \pm 0.006), \\
  \log H_0 & = & 0.2\,M^{\rm corr}_{\rm V} + (5.663 \pm 0.006), 
           \quad \mbox{and} \\
  \log H_0 & = & 0.2\,M^{\rm corr}_{\rm I} + (5.610 \pm 0.007). 
\end{eqnarray}

Here only the absolute magnitudes $M_i$ of the calibrators in
Table~1 must be inserted to obtain $H_0$. But the
calibrators must also be corrected by equations (1$-$3), giving
$M^{\rm corr}_{\rm B}=-19.50\pm0.08$, $M^{\rm corr}_{\rm
  V}=-19.50\pm0.07$, and $M^{\rm corr}_{\rm I}=-19.21\pm0.10$.
Inserting these values into equations (7$-$9) yields
$H_0(B)=57.5\pm2.9$, $H_0(V)=57.9\pm2.7$, and $H_0(I)=58.6\pm3.6$. The
weighted mean of these three values is
\begin{equation}
  \label{eq:10}
  H_0 = 57.9 \pm 1.7 \quad \mbox{(statistical error)}.
\end{equation}
The external errors will be discussed in Section~6.

   Had we not corrected the sample of distant SNe\,Ia and the
calibrators by equations (1$-$3), the result would have been
$H_0=55.0\pm1.6$. The reason for the lower value is that the
calibrators must have Cepheids in their parent galaxies, which hence
are of late type, and SNe\,Ia in late-type galaxies are brighter (and
have lower $\Delta m_{15}$) than their counterparts in E/S0's. The
corrections for the decline rate have sometimes been exaggerated in
the past (Phillips 1993; Hamuy et~al.\ 1996; Suntzeff et~al.\ 1999;
Gibson et~al.\ 1999) leading to larger values of $H_0$. However these
larger decline rate corrections rest on absolute magnitudes from
unreliable distance indicators or on residuals from the Hubble line
where peculiar motions are still not negligible. The corrections in
equations~(1$-$3), based only on distant SNe\,Ia with $v>10\,000\kms$
remove all luminosity dependence of the SNe\,Ia on $\Delta m_{15}$ and
$(B-V)$. Any larger $\Delta m_{15}$ corrections would introduce a
luminosity dependence of opposite sign.

   It may be noted that the coefficients of the colour term in
equations ($1-3$) are significantly smaller than the ones expected
from a standard absorption/reddening ratio. It is therefore likely
that the dependence of luminosity on colour is due to an intrinsic
effect. The colour corrections decrease significantly the scatter
about the Hubble line. They do not influence, however, the value of
$H_0$ because the calibrators and the 36 sample SNe\,Ia have closely
the same colour.

   The value of $H_0$ in equation\,(10) holds only for SNe\,Ia with
$v>10\,000\kms$, i.e. over very large scales. The 17 SNe\,Ia with
$1000\kms<v<10\,000\kms$ give instead a mean value of
$H_0=61.1\pm2.1$, suggesting a local, yet rather large volume of lower
mean density and with a slight overexpansion rate of $\sim\!5\%$ (Tammann
1998a; Zehavi et~al.\ 1998). Combining the evidence from $B$, $V$, and
$I$ it is at present a 2$-$3$\sigma$ effect. Newly discovered SNe\,Ia will
improve on the variance of the expansion field.

\section{{\boldmath $H_0$} from External Evidence}
The available space does not allow a thorough comparison of $H_0$ in
equation~10 with independent evidence. Therefore only an overview of
other distance scales, leading mainly through the Virgo cluster, is
given in Fig.~2. 
The reader finds a more detailed discussion in
Tammann, Sandage, \& Reindl (1999). A compilation of present results
of physical methods, i.e. the Sunyaev-Zeldovich effect, gravitational
lenses, and CMB fluctuations, can be found in, e.g., Tammann (1999;
see also Lasenby et~al.\ 1999). About 75\% of the distance scale in
Fig.~2 rests on the local calibration through Cepheids. It is
therefore not truly independent of the SNe\,Ia.

\begin{figure}[htb]
\begin{center}
{\scriptsize \sf
\epsfxsize=3.18cm   
\setlength{\unitlength}{1mm}
\def\xx{120}
\def\yy{151}
\begin{picture}(\xx,\yy)(0,0)
 \put(0,0){\line(\xx,0){\xx}}
 \put(0,0){\line(0,\yy){\yy}}
 \put(0,\yy){\line(\xx,0){\xx}}
 \put(\xx,0){\line(0,\yy){\yy}}
 \put(110,146){\makebox(0,0)[t]{\normalsize \bf
    H\boldmath$_0$ $\approx$ 58}}
 \put(53,146){\makebox(0,0)[t]{\normalsize \bf H\boldmath$_0$ $=$ 56$\pm$4}}
 \put(53,138.5){\vector(0,0){3.5}}
 \put(40.5,134){\fbox{\parbox{2.2cm}{\centering
    Clusters out to \\ 10\,000\kms}}}
 \put(36,91){\fbox{\epsfbox{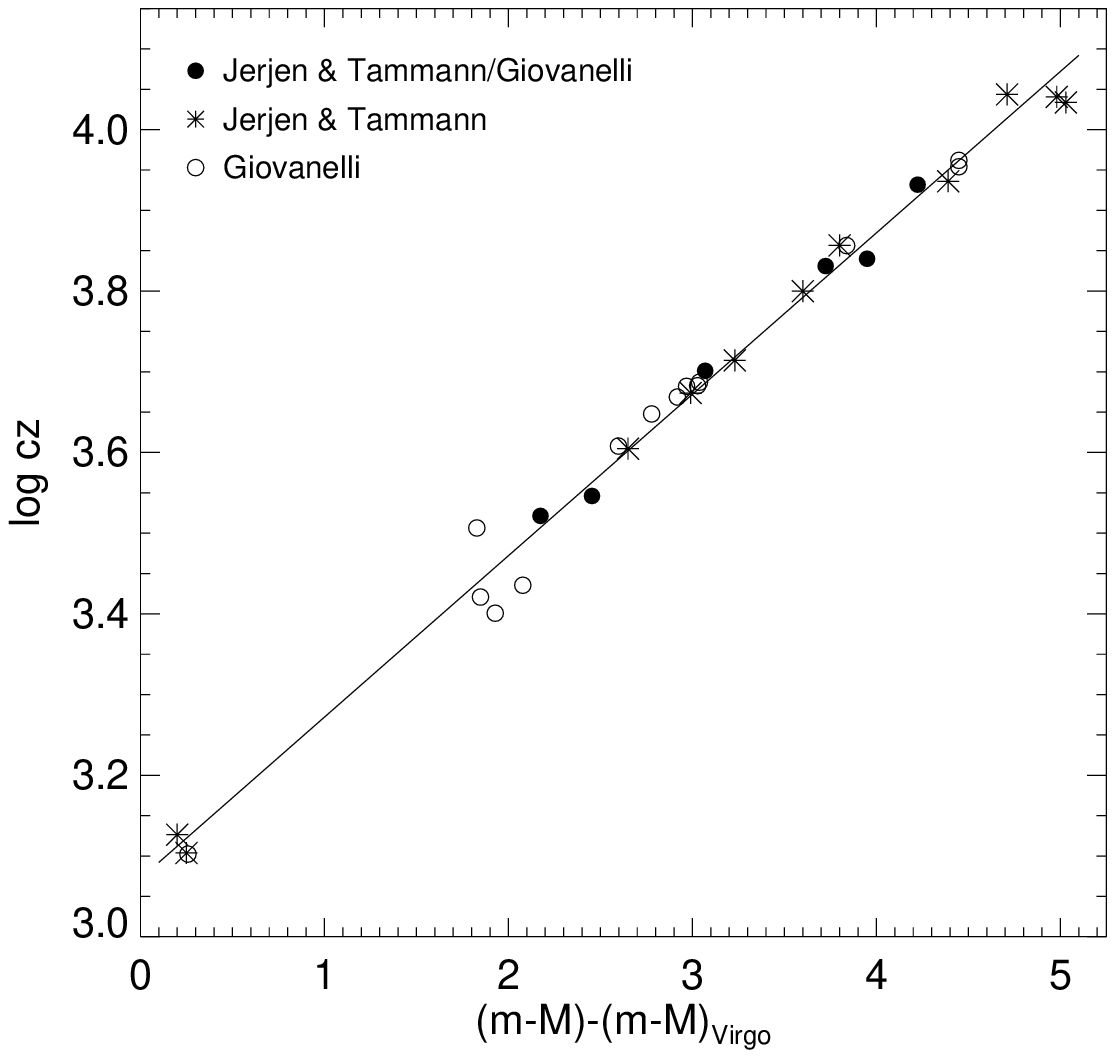}}}
 \put(53,121){\vector(0,0){10}}
 \put(53,80.3){\vector(0,0){9.5}}
 \put(36,74.9){\vector(-3,2){15}}
 \put(70.4,72){\vector(1,1){7.4}}
 \put(16,104){\makebox(0,0)[t]{\normalsize \bf
    (H\boldmath$_0$ $=$ 60$\pm$6)}}
 \put(16,96.5){\vector(0,0){3.5}}
 \put(1,90){\fbox{\parbox{2.6cm}{\centering
    Coma Cluster \\ (m-M)$=$35.29$\pm$0.11 \\
     114$\pm$6$\;$Mpc}}}
 \put(36,72){\fboxrule=1pt \fbox{\parbox{3.2cm}{\centering \vspace*{5pt}
    Virgo Cluster \\  (m-M)$=$31.60$\pm$0.09 \\
    20.9$\pm$0.9$\;$Mpc \vspace*{5pt}}}}
 \put(92,92){\makebox(0,0)[t]{\normalsize \bf (H\boldmath$_0$ $=$
   58$\pm$10)}}
 \put(92,84.5){\vector(0,0){3.5}}
 \put(78,78){\fbox{\parbox{2.6cm}{\centering
   Fornax Cluster \\  (m-M)$=$31.80$\pm$0.09 \\
   22.9$\pm$1.0$\;$Mpc}}}
 \put(58,56){\fbox{\parbox{0.8cm}{\centering
   Leo \\ Group}}}
 \put(62.5,61){\vector(-1,4){1}}
%
\put(14,56.7){\vector(3,2){22}}
\put(28,57.4){\vector(1,1){8}}
\put(40,58){\vector(1,3){2.3}}
\put(75.3,57.5){\vector(-2,3){5}}
\def\py{51}
 \put(10,58){\vector(0,0){27}}
 \put(10,58){\vector(0,0){27}}
\put(10,\py){\circle{40}}
\put(10,\py){\makebox(0,0)[c]{\parbox{1cm}{\centering \scriptsize
      Globular \\ clusters}}}
\put(25,\py){\circle{40}}
\put(25,\py){\makebox(0,0)[c]{\parbox{1cm}{\centering \scriptsize
      Novae}}}
\put(40,\py){\circle{40}}
\put(40,\py){\makebox(0,0)[c]{\parbox{1cm}{\centering \scriptsize
      D$_{\rm n}-\sigma$ \\ relation}}}
\put(25,36.5){\vector(0,0){7.5}}
\put(22,36.5){\vector(-1,1){8.3}}
\put(28,36.5){\vector(1,1){8.3}}
\put(78,\py){\circle{40}}
\put(78,\py){\makebox(0,0)[c]{\parbox{1cm}{\centering \scriptsize
      TF \\ relation}}}

\put(93,56.7){\vector(-3,2){22.3}}
\put(97,58){\vector(0,0){15.0}}
\put(97,\py){\circle{40}}
\put(97,\py){\makebox(0,0)[c]{\parbox{1cm}{\centering \scriptsize
      SNe\,Ia}}}
\def\py{30}
 \put(15,\py){\fbox{\parbox[c]{2.6cm}{\centering
   M\,31 \\ (m-M)$=$24.44$\pm$0.15 \\
     773$\pm$55$\;$kpc}}}
 \put(68,28){\fbox{\parbox{2.6cm}{\centering
   LMC \\ (m-M)$=$18.50$\pm$0.04 \\
     50.1$\pm$1.0$\;$kpc}}}
%
\def\py{10}
 \put(10,17){\vector(0,0){27}}
\put(15.8,14.4){\vector(4,1){52}}
\put(19.8,15.4){\vector(-4,-1){4}}
\put(10,\py){\circle{60}}
\put(10,\py){\makebox(0,0)[c]{\parbox{1cm}{\centering \scriptsize
      RR\,Lyr \\ stars}}}
%
\def\dx{53}
 \put(\dx,64){\vector(0,0){1}}
 \put(\dx,62){\circle*{0.5}}
 \put(\dx,60){\circle*{0.5}}
 \put(\dx,58){\circle*{0.5}}
 \put(\dx,56){\circle*{0.5}}
 \put(\dx,54){\circle*{0.5}}
 \put(\dx,52){\circle*{0.5}}
 \put(\dx,50){\circle*{0.5}}
 \put(\dx,48){\circle*{0.5}}
 \put(\dx,46){\circle*{0.5}}
 \put(\dx,44){\circle*{0.5}}
 \put(\dx,42){\circle*{0.5}}
 \put(\dx,40){\circle*{0.5}}
 \put(\dx,38){\circle*{0.5}}
 \put(\dx,36){\circle*{0.5}}
 \put(\dx,34){\circle*{0.5}}
 \put(\dx,32){\circle*{0.5}}
 \put(\dx,30){\circle*{0.5}}
 \put(\dx,28){\circle*{0.5}}
 \put(\dx,26){\circle*{0.5}}
 \put(\dx,24){\circle*{0.5}}
 \put(\dx,22){\circle*{0.5}}
 \put(\dx,20){\circle*{0.5}}
 \put(\dx,18){\circle*{0.5}}
\put(47,14){\vector(-3,2){16}}
\put(59,14){\vector(3,2){13}}
\put(60.5,15){\vector(-3,-2){1.5}}
\put(54,17){\vector(2,3){19}}
\put(53.4,17){\vector(1,4){8.9}}
\put(57.2,16){\line(4,3){10}}
\put(82.8,35.2){\vector(4,3){12}}

\put(\dx,\py){\circle{60}}
\put(\dx,\py){\makebox(0,0)[c]{\scriptsize Cepheids}}
 \put(98.6,41.1){\vector(-1,3){1}}
 \put(98.5,41.2){\circle*{0.5}}
 \put(99.1,39.4){\circle*{0.5}}
 \put(99.7,37.6){\circle*{0.5}}
 \put(100.3,35.8){\circle*{0.5}}
 \put(100.9,34.0){\circle*{0.5}}
 \put(101.5,32.2){\circle*{0.5}}
 \put(102.1,30.4){\circle*{0.5}}
 \put(102.7,28.6){\circle*{0.5}}
 \put(103.3,26.8){\circle*{0.5}}
 \put(103.9,25){\circle*{0.5}}
 \put(104.5,23.2){\circle*{0.5}}
 \put(105.1,21.4){\circle*{0.5}}
 \put(105.7,19.8){\circle*{0.5}}
 \put(106.3,18){\circle*{0.5}}
 \put(110,17){\vector(0,0){124}}
 \put(104,14){\vector(-3,2){13}}
\put(110,\py){\circle{60}}
\put(110,\py){\makebox(0,0)[c]{\parbox{1cm}{\centering \scriptsize
      Physical \\ methods}}}
%
\end{picture}
}
  \end{center}
  \caption{Schematical presentation of the distance scale based on
    various determinations of the Virgo cluster distance.} 
  \label{fig:schematic_Ho_all}
\end{figure}
\clearpage

   The route to $H_0$ through field galaxies is not satisfactory,
because it will never reach out to $10\,000\kms$ and beyond. In
addition the route is technically difficult because of selection
effects, causing the galaxies which are missed in a sample to be
almost as important as those which constitute the sample. 
In spite of this, careful treatment of the Malmquist bias leads to
values of $H_0=55-60$ at distances of $1000-5000\kms$ 
(Sandage 1996a,b, 1999; Theureau et~al.\ 1997; Goodwin et~al.\ 1997;
Federspiel 1999). -- Values as high as $H_0=70$ are an unfailing
indication that the bias problem has been underestimated. Only too
easily a key project for the determination of $H_0$ turns this way
into the key to a Hubblegate.

\section{Conclusions}
The 17 galaxies with known Cepheid distances outside the Local Group
and outside the Virgo cluster yield $H_0=65\pm4$. But reaching out to
barely $1100\kms$ this value has no cosmological significance. It is
therefore mandatory to extend the distance scale.

   By far the most reliable way to extend the distance scale is by
SNe\,Ia which are, after standardisation as to decline rate and color,
unparalleled standard candles. The number of Cepheid-calibrated and distant
SNe\,Ia out to $30\,000\kms$ is now sufficiently large to make the
statistical error of $H_0$ negligible (cf. equation~7).

Sources of systematic errors are: (1) the adopted zeropoint of
$(m-M)_{\rm LMC}=18.50$, a value which is likely to be too small by
$0\mag06\pm0\mag10$. This will reduce $H_0$ by $3\pm5\%$. (2) The
selection effect against faint Cepheids at the detection limit
(Sandage 1988) may systematically underestimetethe distances in
Table~1 by $5\pm5\%$ with a corresponding reduction of $H_0$. (3)
Remaining small metallicity effects of the P-L relation will probably
not change the calibration by more than $0\mag06$ (3\%). (4) The
coefficients of the $\Delta m_{15}$-term in equations (1$-$3) carry a
statistical error of 0.2; increasing (decreasing) the coefficients by
so much increases (decreases) $H_0$ by 2.7 units (5\%). (5) The nearly
perfect agreement of the mean color of the calibrators and the
obviously unreddened SNe\,Ia on the one hand and the remaining distant
SNe\,Ia on the other hand (cf. Section~4) make it improbable that the
latter still carry an unaccounted mean reddening of more than
$0\mag02$; about half of the corresponding absorption is automatically
corrected for by the color term in equations (1$-$3); the other half
can increase $H_0$ by 2\% at most.

   Compounding the systematic errors and adding them to equation~(7)
leads to a most probable value of
\begin{displaymath}
   H_0=57.9^{+4.2}_{-8.3}.
\end{displaymath} 

\bigskip

\acknowledgements
The authors thank the Swiss National Science Foundation for financial
support.


\clearpage
\begin{references}
\reference Alibert, Y., Baraffe, I., Hauschildt, P., \& Allard, F. 1999,
  \aap, 344, 551 
\reference Baraffe, I., Alibert, Y., M{\'e}ra, D., Chabrier, G., \&
  Beaulieu, J.-P. 1998, \apj, 499, L205
\reference Bono, G., Marconi, M. \& Stellingwerf, R.F. 1998, 
  \apjs, 122, 167
\reference Branch, D. 1998, \araa, 36, 17
\reference Branch, D., Fisher, A., \& Nugent, P. 1993, \aj, 106, 
   2383
\reference Chaboyer, B. 1999, private communication
\reference Della Valle, M., Kissler-Patig, M., Danziger, J., \& Storm, J.
  1998, \mnras, 299, 267
\reference Di Benedetto, G.P. 1997, \apj, 486, 60
\reference Feast, M. 1999, \pasp, 111, 775
\reference Feast, M., \& Catchpole, R.M. 1997, \mnras, 286, L1
\reference Federspiel, M. 1999, Ph.D. Thesis, Univ. of Basel
\reference Federspiel, M., Tammann, G.A., \& Sandage, A. 1998, \apj,
  495, 115 
\reference Ferrarese, L., et~al. 1996, \apj, 464, 568
\reference Freedman, W.L. 1999, ,\baas, 194, 3909
\reference Gibson, B.K. 1999, \apj, in press, astro-ph/99\,08\,149 
\reference Gilmozzi, R., \& Panagia, N. 1999, Space Telescope Science
  Inst. Preprint Series No. 1319
\reference Goodwin, S.P., Gribbin, J., \& Hendry, M.A. 1997, \aj, 114, 2212
\reference Gratton, R. 1998, in 19th Texas Symposium on
    Relativistic Astrophysics and Cosmology, in press
\reference  Hamuy, M., Phillips, M.M., Maza, J., Suntzeff, N.B., Schommer,
  R.A., \& Aviles, R. 1996, \aj, 112, 2408
\reference H{\"o}flich, P., \& Khokhlov, A. 1996, \apj, 457, 500
\reference Kennicutt, R.C., Mould, R.J., \& Freedman, W.L. 1998,
  preprint
\reference Kraan-Korteweg, R. 1986, \aaps, 66, 255 
\reference Lasenby, A., Bridle, S., Hobson, M., \& Efstathiou,
G. 1999, in  19th Texas Symposium on Relativistic Astrophysics and
Cosmology, preprint
\reference Madore, B.F., \& Freedman, W.L. 1991, \pasp, 103, 933
\reference Mazumdar, A. 1999, this volume 
\reference Mould, J.R., et~al. 1999, \apj, in press, 
  astro-ph/99\,09\,260 
\reference Narasimha, D., \& Mazumdar, A. 1998, astro-ph/98\,03\,195 
\reference Parodi, B.R., Saha, A., Sandage, A., \& Tammann, G.A. 1999,
  \apj, submitted
\reference Phillips, M.M. 1993, \apj, 413, L105
\reference Saha, A., Labhardt, L., Schwengeler, H., Macchetto, F.D., 
  Panagia, N., Sandage, A., \& Tammann, G.A. 1994, \apj, 425, 14 
\reference Saha, A., Sandage, A., Labhardt, L., Schwengeler, H., Tammann,
  G.A., Panagia, N., \& Macchetto, F.D. 1995, \apj, 438, 8 
\reference Saha, A., Sandage, A., Labhardt, L., Tammann, G.A., 
  Macchetto, F.D., \&  Panagia, N. 1996a, \apj, 466, 55 
\reference Saha, A., Sandage, A., Labhardt, L., Tammann, G.A., 
  Macchetto, F.D., \&  Panagia, N. 1996b, \apjs 107, 693 
\reference Saha, A., Sandage, A., Labhardt, L., Tammann, G.A., 
  Macchetto, F.D., \&  Panagia, N. 1997, \apj, 486, 1 
\reference Saha, A., Sandage, A., Labhardt, L., Tammann, G.A., Macchetto,
  F.D., \& Panagia, N. 1999, \apj, 522, 802 
\reference Saio, H., \& Gautschy, A. 1998, \apj, 498, 360 
\reference Sandage, A. 1988, \pasp, 100, 935 
\reference Sandage, A. 1996a,b, \aj, 111, 1 \& 18
\reference Sandage, A., 1999, preprint
\reference Sandage, A., \& Bell, R.A., \& Tripicco, M.J. 1998,
  \apj, 522, 250 
\reference Sandage, A., \& Tammann, G.A. 1968, \apj, 151, 531 
\reference Sandage, A., \& Tammann, G.A. 1971, \apj, 167, 293 
\reference Sandage, A., \& Tammann, G.A. 1998, \mnras, 293, L23 
\reference Schlegel, D., Finkbeiner, D., \& Davis, M. 1998, \apj, 500,
   525.
\reference Smoot, G., et~al. 1992, \apj, 396, L1 
\reference Sonneborn, G., et~al. 1997, \apj, 477, 848 
\reference Stetson, P.B., et~al. 1998, \apj, 508, 491 
\reference Suntzeff, N.B., et~al. 1999, \aj, 117, 1175
\reference Tammann, G.A. 1997, J. Astrophys. Astro., 18, 271
\reference Tammann, G.A. 1998a, in General Relativity, 8th
    Marcel Grossmann Symp., ed. T. Piran (Singapore: World
    Scientific), p.~243
\reference Tammann, G.A. 1998b, in Harmonizing the Cosmic
    Distance Scale, eds. D. Egret and A. Heck, in press 
\reference Tammann, G.A. 1999, in Dark Matter in Astrophysics and
    Particle Physics, eds. H.V. Klapdor-Kleingrothaus \& L. Baudis
    (Bristol: Inst. Physics Publ.), p.~153, preprint
\reference Tammann, G.A., Sandage, A., \& Reindl, B. 1999, in 19th
  Texas Symposium on Relativistic Astrophysics and Cosmology, preprint,
  astro-ph/99\,04\,360 
\reference Tanvir, N.R., Shanks, T., Ferguson, H.C., \& Robinson, D.R.T.
  1995, Nature, 377, 27  
\reference Theureau, G., et~al. 1997, \aap, 322, 730
\reference Thim, F. 1999, this volume 
\reference Turner, A., et~al. 1998, \apj, 505, 207 
\reference Walker, A.R. 1999a, in Post-Hipparcos Cosmic Candles,
  eds. A. Heck \& F. Caputo (Dordrecht: Kluwer), p.~125 
\reference Walker, A.R. 1999b, private communication 
\reference Wang, L., H{\"o}flich, P., \& Wheeler, J.C. 1997, \apj,
  483, L29 
\reference Zehavi, I., Riess, A.G., Kirshner, R.P., \& Dekel, A. 1998,
  \apj, 503, 483 
\end{references}
\end{document}